\documentclass[twocolumn,aip,nofootinbib,jcp,reprint,amsmath,amssymb]{revtex4-1}

\usepackage{natbib,hyperref}
\usepackage{amsmath}
\usepackage{graphicx}
\usepackage{hyperref} 
\usepackage{dcolumn}
\usepackage{bm}
\usepackage{epsfig,color,xspace,multirow,xr,bbold}
\usepackage[all]{xy}
\usepackage{setspace}
\usepackage{url}
\usepackage{soul}
\usepackage[colorinlistoftodos]{todonotes}
\usepackage{longtable}
\usepackage{tabularx}
\usepackage{adjustbox}
\usepackage{threeparttable} 
\usepackage{natbib,hyperref}
\usepackage{float}
\usepackage{placeins}
\usepackage[utf8]{inputenc}
\usepackage[T1]{fontenc}
\usepackage{soul} 

\usepackage[normalem]{ulem}
\usepackage{xcolor}
\newcommand{\RRef}[1]{Ref.~\onlinecite{#1}}
\newcommand{\RRefs}[1]{Refs.~\onlinecite{#1}}
\newcommand\notsotiny{\@setfontsize\notsotiny\@vipt\@viipt}
\usepackage{soul} 
\newcommand{\orcid}[1]{\href{https://orcid.org/#1}{\includegraphics[width=8pt]{orcid.png}}}
\begin{document}

\title{
Enhancing NMR Shielding Predictions of Atoms-in-Molecules Machine Learning Models 
with Neighborhood-Informed Representations
}
\date{\today}

\author{
Surajit Das, 
Raghunathan Ramakrishnan 
}

\email{ramakrishnan@tifrh.res.in}
\affiliation{$^1$Tata Institute of Fundamental Research, Hyderabad 500046, India}


\begin{abstract}
\noindent
Accurate prediction of nuclear magnetic resonance (NMR) shielding with machine learning (ML) models remains a central challenge for data-driven spectroscopy.
We present atomic variants of the Coulomb matrix (aCM) and bag-of-bonds (aBoB) descriptors, and extend them using radial basis functions (RBFs) to yield smooth, per-atom representations (aCM-RBF, aBoB-RBF). Local structural information is incorporated by augmenting each atomic descriptor with contributions from the $n$ nearest neighbors, resulting in the family of descriptors, aCM-RBF($n$) and aBoB-RBF($n$).
For $^{13}$C shielding prediction on the QM9NMR dataset (831,925 shielding values across 130,831 molecules), aBoB-RBF(4) achieves an out-of-sample mean error of 1.69 ppm, outperforming models reported in previous studies.
While explicit three-body descriptors further reduce errors at a higher cost, aBoB-RBF(4) offers the best balance of accuracy and efficiency.
Benchmarking on external datasets comprising larger molecules (GDBm, Drug12/Drug40, and pyrimidinone derivatives) confirms the robustness and transferability of aBoB-RBF(4), establishing it as a practical tool for ML-based NMR shielding prediction.
\end{abstract}

\maketitle

\section{Introduction}\label{sec:introduction}
Nuclear magnetic resonance (NMR) spectroscopy is a versatile, non-destructive technique for elucidating the three-dimensional structures of molecules and crystals.
Ab initio NMR calculations~\cite{helgaker1999ab,mulder2010nmr}, typically based on density functional theory (DFT), complement experiments by validating proposed molecular structures~\cite{nicolaou2005chasing,suyama2011survey,maier2009structural}, assigning chemical shifts~\cite{grimblat2016computational}, resolving cases where experimental spectra are insufficient~\cite{das2020metabolite,cuadrado2024computationally}, and determining stereochemistry in diastereomers~\cite{smith2009assigning,smith2010assigning}.
However, the broad conformational diversity of large molecules often limits the scope of such ab initio approaches, as the cost of accurate geometry optimizations and NMR parameter calculations is prohibitive for large-scale, high-throughput screening required in areas such as drug discovery and metabolite annotation.

In contrast to DFT-based modeling, empirical methods such as those in ChemDraw~\cite{chemnmr} and MarvinSketch~\cite{csizmadia1999marvinsketch} predict NMR chemical shifts directly from atomic connectivities, without requiring three-dimensional structures.
ChemDraw employs an additive model with thousands of parameters fitted to the curated ChemNMR dataset~\cite{chemnmr}, yielding typical prediction errors of 3.8 ppm for $^{13}$C shifts across several environments and 0.2–0.3 ppm for $^{1}$H shifts of CH$x$ groups (with C connected to 1--3 H atoms)~\cite{cheeseman2000predicting}.
MarvinSketch relies on the hierarchical ordered spherical description of environment (HOSE)~\cite{bremser1978hose}, which is trained on the NMRShiftDB database~\cite{steinbeck2004nmrshiftdb}.
While such empirical approaches offer near-instantaneous predictions and can distinguish many common isomers, they frequently break down for under-represented or complex environments that occur across the vast chemical space, such as those enumerated in the GDB17 database~\cite{ruddigkeit2012enumeration} of small organic molecules with up to 17 CONF atoms.

Machine learning (ML) provides a powerful data-driven alternative for modeling molecular properties~\cite{rupp2012fast}. Once trained on first-principles calculations or other reference data, ML models deliver predictions with near-empirical speed while retaining the accuracy of the underlying target method~\cite{ramakrishnan2017machine}. A growing body of work~\cite{rupp2015machine,gupta2021revving,zou2023deep,shiota2024universal,xu2025toward,ramos2025interplay,gaumard2022regression,ondar2023predicting,el2025amide,paruzzo2018chemical,gerrard2020impression,buning2024machine,howarth2020dp4,wang2025nmrextractor,lemm2024impact,cordova2022machine,cordova2023atomic,unzueta2025predicting,steinbeck2004nmrshiftdb,yuan2025qme14s} has demonstrated the applicability of ML combined with quantum chemistry for predicting a wide range of NMR properties in molecules and solids.
To support such models, several dedicated datasets have been developed and curated, including QM9NMR~\cite{gupta2021revving}, IMPRESSION~\cite{gerrard2020impression}, NMRShiftDB~\cite{steinbeck2004nmrshiftdb}, and QMe14S~\cite{yuan2025qme14s}, which provide thousands of examples required to reduce training set bias. Among these, QM9NMR~\cite{gupta2021revving} is one of the largest, containing structures and NMR shielding parameters for $^{1}$H, $^{13}$C, $^{15}$N, $^{17}$O, and $^{19}$F in 130,831 molecules from the QM9 dataset~\cite{ramakrishnan2014quantum} of small organic molecules with up to 9 CONF atoms, including 831,925 $^{13}$C shieldings. This dataset has enabled the training of ML models with up to 100,000 (i.e., 100k) examples.
Gupta et al.~\cite{gupta2021revving} trained a kernel ridge regression (KRR) model on 100k QM9NMR entries using the  Faber--Christensen--Huang--Lilienfeld (FCHL) many-body descriptor~\cite{faber2018alchemical}, achieving an out-of-sample error of 1.88 ppm for $^{13}$C shifts of 50k hold-out atoms. Using the same split, Shiota et al.~\cite{shiota2024universal} reported a similar error of 1.87 ppm with a KRR model based on a graph neural network–derived transfer learning embedding and the MACE-OFF23-small descriptor (the ``small'' variant of a transferable organic force field derived from the message passing atomic cluster expansion architecture~\cite{kovacs2025mace}). Notably, when applied to drug molecules outside the training domain, the model of Shiota et al. delivered smaller errors than that of Gupta et al.

The objective of this study is to develop systematically improved atom-centered structural descriptors for predicting NMR chemical shifts, trained on the QM9NMR dataset. We introduce atomic variants of the Coulomb matrix (CM)~\cite{rupp2012fast} and bag-of-bonds (BoB)~\cite{hansen2015machine} descriptors in both discrete and continuous formulations. To further capture local chemical environments, we augment these atomic descriptors with information on nearest neighbor atoms. The resulting representations achieve lower out-of-sample errors than existing approaches, establishing new benchmarks for diverse classes of organic molecules.
The remainder of this article is organized as follows.
Section~\ref{sec:dataset} introduces the datasets used in this study and outlines the computational details.
Section~\ref{sec:ml} describes the kernel ridge regression (KRR) approach and the construction of the atomic descriptors.
Section~\ref{sec:results} presents the optimization of hyperparameters, evaluates the efficiency of the descriptors, and discusses model performance for predicting $^{13}$C chemical shifts across different datasets.

\section{Datasets and Computational Details}\label{sec:dataset}
We collected QM9NMR, Drug12, Drug40, and GDBm datasets from the  QM9NMR dataset's repository\cite{qm9nmr} developed in a previous study\cite{gupta2021revving}.  QM9 molecules were optimized using the B3LYP/6-31G(2{\it df,p}) DFT/basis set combination. For ML modeling, the target quantities used are NMR parameters of QM9 molecules, calculated at the mPW1PW91/6-311+G(2{\it d,p}) level, using geometries in QM9 determined with the B3LYP/6-31G(2{\it df,p}) method. 
These values were collected from the QM9NMR dataset\cite{qm9nmr}. 
In this study, we trained ML models using descriptors based on minimum energy geometries calculated with the PM7 semi-empirical model, as a previous study has shown the efficiency of using descriptors derived from geometries determined with inexpensive methods\cite{ramakrishnan2015big}. 
For QM9NMR, Drug12, Drug40, and GDBm datasets, PM7 geometries were collected from the QM9NMR dataset\cite{qm9nmr}.
We computed NMR parameters for 208 biologically relevant pyrimidinone molecules using the same protocol employed in the QM9NMR dataset~\cite{gupta2021revving}. Specifically, NMR properties were calculated at the mPW1PW91/6-311+G(2\textit{d,p}) level of theory, using minimum-energy geometries obtained with B3LYP/6-31G(2\textit{df,p}) from the QM9 dataset~\cite{ramakrishnan2014quantum}.
Table~\ref{tab:dataset} summarizes the molecular datasets explored in this study, with specific details for each dataset provided in the subsections below.

\begin{table}[h]
    \centering
    \caption{Summary of datasets explored in this study: `min-CONF' and `max-CONF' are the minimum and maximum number of CONF atoms present in a molecule of that dataset. } 
    \label{tab:dataset}
    \begin{threeparttable}
        \begin{tabular}{l rr r r }
            \hline
            Dataset & Molecules &&  min-CONF & max-CONF  \\
            \hline
            QM9            & 130831 && 1    & 9     \\
            Drug12         & 12     && 17   & 23    \\
            Drug40         & 40     && 7    & 17    \\
            pyrimidinone   & 208    && 7    & 10    \\
            GBDm (m=10--17)& 200    && 10   & 17    \\
            \hline
        \end{tabular}
    \end{threeparttable}
\end{table}

In this study, we focus on the gas-phase $\sigma_{\rm iso}(^{13}$C)  values (in ppm) as the target property for assessing descriptor performance in ML modeling. Here, $\sigma_{\rm iso}$ denotes the isotropic average (trace) of the shielding tensor. From the isotropic shielding values, $\sigma_{\rm iso}(^{13}{\rm C})$, the chemical shifts are obtained as
\begin{eqnarray}
\delta(^{13}{\rm C}) = \sigma_{\rm iso}^{\rm TMS}(^{13}{\rm C}) - \sigma_{\rm iso}(^{13}{\rm C}),
\label{eq:chemicalshift}
\end{eqnarray}
where tetramethylsilane (TMS) is used as the standard reference for $^{13}$C NMR.
The gas-phase $\sigma_{\rm iso}(^{13}{\rm C})$ of TMS, calculated at the mPW1PW91/6-311+G(2{\it d,p}) level using the minimum-energy geometry optimized with B3LYP/6-31G(2{\it df,p}), was taken from the QM9NMR repository and used as the reference value, $\sigma_{\rm iso}^{\rm TMS} = 186.9704$ ppm~\cite{qm9nmr}.

For the $\Delta$-ML modeling~\cite{ramakrishnan2015big}, baseline NMR calculations were carried out at the HF/STO-3G level using PM7-optimized geometries obtained from the QM9NMR repository~\cite{qm9nmr}. The HF baseline was originally chosen because it is conceptually simple, implementation-independent compared to DFT methods, and widely available across quantum chemistry software packages. 
To further explore alternatives, we also evaluated the cost of using the mPW1PW91 functional with the minimal basis set, STO-3G for a representative molecule containing 9 CONF atoms (SMILES: {\tt CC1CCCCC1C=O}), NMR shielding tensor calculations at the mPW1PW91/6-311+G(2\textit{d,p}) and mPW1PW91/STO-3G levels require 776.5 and 21.6 seconds of CPU time, respectively, amounting to a speedup of roughly 35-fold. This substantial reduction in computational cost highlights the suitability of minimal-basis calculations as baselines for $\Delta$-ML modeling of NMR chemical shifts.

Baseline NMR calculations for all QM9 molecules were newly performed at the mPW1PW91/STO-3G level using Gaussian16~\cite{frisch2016gaussian} and the same PM7-level geometries obtained from the QM9NMR repository~\cite{qm9nmr}. Since mPW1PW91 is also the functional used to generate the target NMR properties, the resulting $\Delta$-ML model learns to correct both (i) the geometric difference between PM7 and B3LYP/6-31G(2\textit{df,p}) structures and (ii) the basis-set change from STO-3G to 6-311+G(2\textit{d,p}).

\subsection{QM9 and QM9NMR}\label{subsec:qm9}
    
The QM9 dataset~\cite{ramakrishnan2014quantum,qm9pack} is a widely used benchmark for ML modeling of molecular properties. 
Building on this resource, the QM9NMR dataset~\cite{gupta2021revving} provides isotropic NMR chemical shielding values ($\sigma_{\rm iso}$) for all nuclei present in QM9: 
1,208,486 $^{1}$H, 
831,925 $^{13}$C, 
132,498 $^{15}$N, 
183,265 $^{17}$O, and 
3,036 $^{19}$F. 
Shieldings are reported for both gas-phase molecules and in implicit-solvent environments (CCl$_4$, THF, acetone, methanol, and DMSO). 

\begin{figure}
    \centering
    \includegraphics[width=1\linewidth]{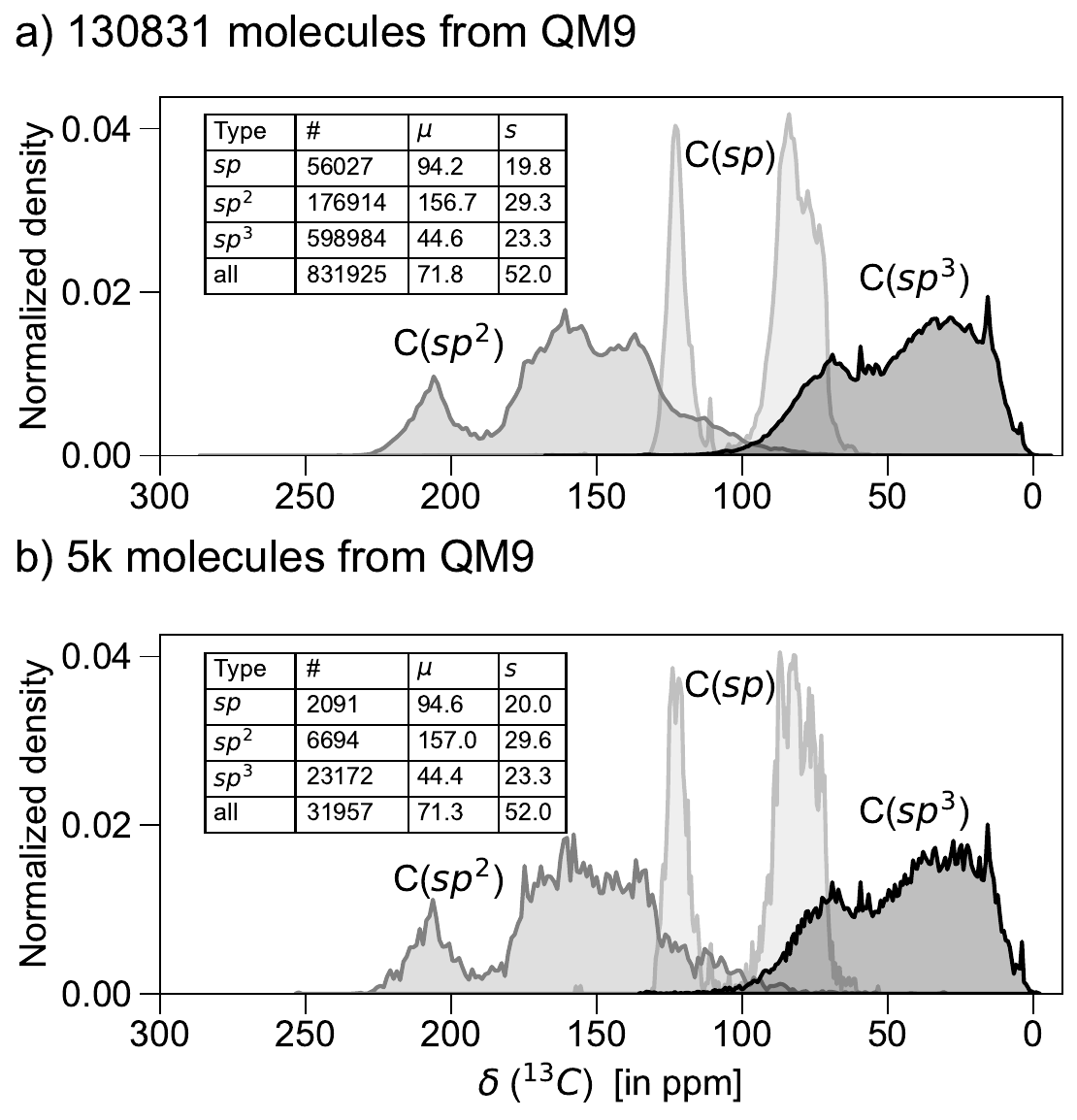}
    \caption{Distribution of $^{13}$C NMR chemical shifts, $\delta(^{13}{\rm C})$, in ppm, in the QM9NMR dataset and its subset with 5k molecules partitioned according to hybridization state of the carbon atom. The table in the inset presents  
    `\#': the number of atoms, while `$\mu$' and `$s$' denote the mean value and the standard deviation of the distribution.}
    \label{fig:qm9_13c_dist}
\end{figure}

Exhaustive benchmarking of ML models on large datasets such as QM9NMR is computationally demanding, as hyperparameter optimization must consider not only the ML ansatz but also parameters in the descriptor definitions.
To make this search tractable, we extract a random subset of 5k molecules from QM9, which is used exclusively for optimization of descriptor parameters.
To assess the representativeness of the 5k subset relative to the full QM9NMR dataset, we compare the distributions of $\delta(^{13}$C) across different hybridization types, as shown in Figure~\ref{fig:qm9_13c_dist}. The two datasets exhibit similar distributions in both chemical shift ranges and relative amplitudes. 
This close agreement indicates that the 5k subset adequately captures the diversity of chemical environments in QM9NMR, justifying its use for rapid parameter optimization before scaling calculations to larger datasets.

In Figure~\ref{fig:qm9_13c_dist}, the $\delta(^{13}{\rm C})$ distribution of the QM9NMR dataset is shown by carbon hybridization. 
The $sp^3$ carbons ($\sim$600k atoms) span a broad range of 0--100 ppm, while the $sp^2$ carbons ($\sim$177k atoms) extend over 100--225 ppm. 
The $sp$ carbons ($\sim$56k atoms) fall in the mid-range, 50-125 ppm, reflecting anisotropic fields associated with triple bonds. 
Their distribution is bimodal, with peaks corresponding to carbons in $-$C\#N and $-$C\#C$-$ groups (where \# denotes a triple bond in SMILES notation~\cite{weininger1988smiles}). 
The right-most peak arises from the deshielded carbon bound to electronegative nitrogen ($-$C\#N), whereas the left-most peak corresponds to the more shielded carbon bound to another carbon ($-$C\#C$-$). 
Because $-$C\#N can extend only in one direction, while $-$C\#C$-$ can extend from both ends, the latter shows a broader distribution.

\subsection{Drug12 and Drug40}
    
For validation of the ML models, two completely out-of-sample datasets, Drug12 (twelve drug molecules from \RRef{corey2007molecules}) and Drug40 (a subset of forty molecules from the GDB17 dataset \RRef{ruddigkeit2012enumeration}), are taken into consideration. These two datasets are also used in previous works \RRefs{gupta2021revving,shiota2024universal}. In the Drug40 dataset, molecules contain 7 to 17 heavy atoms (Table~\ref{tab:dataset}), whereas the training of the ML model is done with at most 9 atoms in the QM9 dataset. In Drug12, the heavy atoms present in a molecule are much higher, 17 to 23. As these datasets contain a larger number of heavy atoms than the QM9 dataset, they are good candidates for the validation of the ML models. 
        
Also, as the heavy atoms are much higher in number in these datasets, they will ensure the transferability of the ML models to large molecular compounds. That is, even though the ML model is trained on a dataset containing a low number of heavy atoms, they are transferable to larger molecules. The PM7 geometries of the drug molecules are obtained from the QM9NMR repository~\cite{qm9nmr}.

\subsection{GDBm}
As a validation set, we also employ the GDBm dataset as a validation set, where $m$ ranges from 10 to 17, denoting the number of heavy atoms in a molecule. For example, GDB13 contains molecules with up to 13 CONF atoms. For each $m$, 25 molecules are selected, yielding a total of 200 molecules in the GDBm set. This dataset was previously used as a validation benchmark in the QM9NMR study~\cite{gupta2021revving}. In the present work, GDBm serves to assess the transferability and accuracy of the proposed descriptors. The PM7-optimized geometries of the GDBm molecules are obtained from the QM9NMR repository~\cite{qm9nmr}.

\subsection{Pyrimidinone}\label{sec:pyrimidinone}
For further validation of the ML models of $\delta(^{13}{\rm C})$ presented in this study, we consider a set of substituted pyrimidinone molecules. 
The dataset comprises 208 biomolecules with substitutions ($-$CH$_3$, $-$NH$_2$, and $-$F), including 16 uracil [pyrimidine-2,4(1H,3H)-dione], 64 pyrimidine-2(1H)-one, 64 pyrimidine-4(1H)-one, and 64 pyrimidine-4(3H)-one derivatives. 
These molecules contain 7--10 heavy atoms, making them suitable candidates for validation. This set was previously used for evaluating ML predictions of core electron binding energies~\cite{tripathy2024chemical}. In the present work, we extend their use to assess NMR chemical shifts, thereby broadening the scope of properties available for these biologically relevant molecules.

For the pyrimidinone dataset, geometries calculated with the universal force field (UFF) were collected from a repository\cite{cebeconf} developed for a previous study\cite{tripathy2024chemical}. 
We optimized these geometries with  B3LYP/6–31G(2{\it df,p}) and calculated  NMR shielding parameters as analytical second derivatives of total energy
with respect to magnetic moments and magnetic field~\cite{gauss2000molecular} with mPW1PW91/6-311+G(2{\it d,p}) using Gaussian16 \cite{frisch2016gaussian}. ML predictions of the chemical shifts of pyrimidinone molecules were done using geometries determined with PM7 using the MOPAC software~\cite{stewart1990mopac}.

\section{Atoms-in-molecules Machine Learning Modeling}\label{sec:ml}

\subsection{Atomic representations of local chemical environments\label{sec:representations}}
In ML modeling of molecules and materials, a descriptor (or representation) provides a mathematical mapping of atomistic structural features. Global descriptors encode the entire molecular structure to predict properties such as total energy~\cite{hansen2015machine}, whereas local (atomic) descriptors capture the structural environment around individual atoms to predict atom-specific properties. The latter approach is commonly referred to as atoms-in-molecules (AIM) machine learning~\cite{rupp2015machine}, which is particularly relevant for NMR shielding modeling. Since NMR shielding is a property of a quasi-atom within a molecule, with the observed chemical shift arising from its local chemical environment, AIM-ML offers a natural framework for such predictions.
More recently, deep learning-based interatomic potential descriptors have been developed to encode atomic environments using graph neural networks (GNNs)~\cite{scarselli2008graph}. For example, the `AtmEnV' descriptor is derived from the SchNet architecture~\cite{schutt2018schnet} to represent the local environment of an atom~\cite{tripathy2024chemical}. Shiota et al.~\cite{shiota2024universal} employed GNN-based descriptors for chemical shift prediction, including MACE~\cite{batatia2022mace}, which is a physics-inspired architecture. In the present study, we focus on simpler structure-based descriptors explicitly derived from ab initio geometries.

\begin{figure*}[!htpb]
    \centering
    \includegraphics[width=1\linewidth]{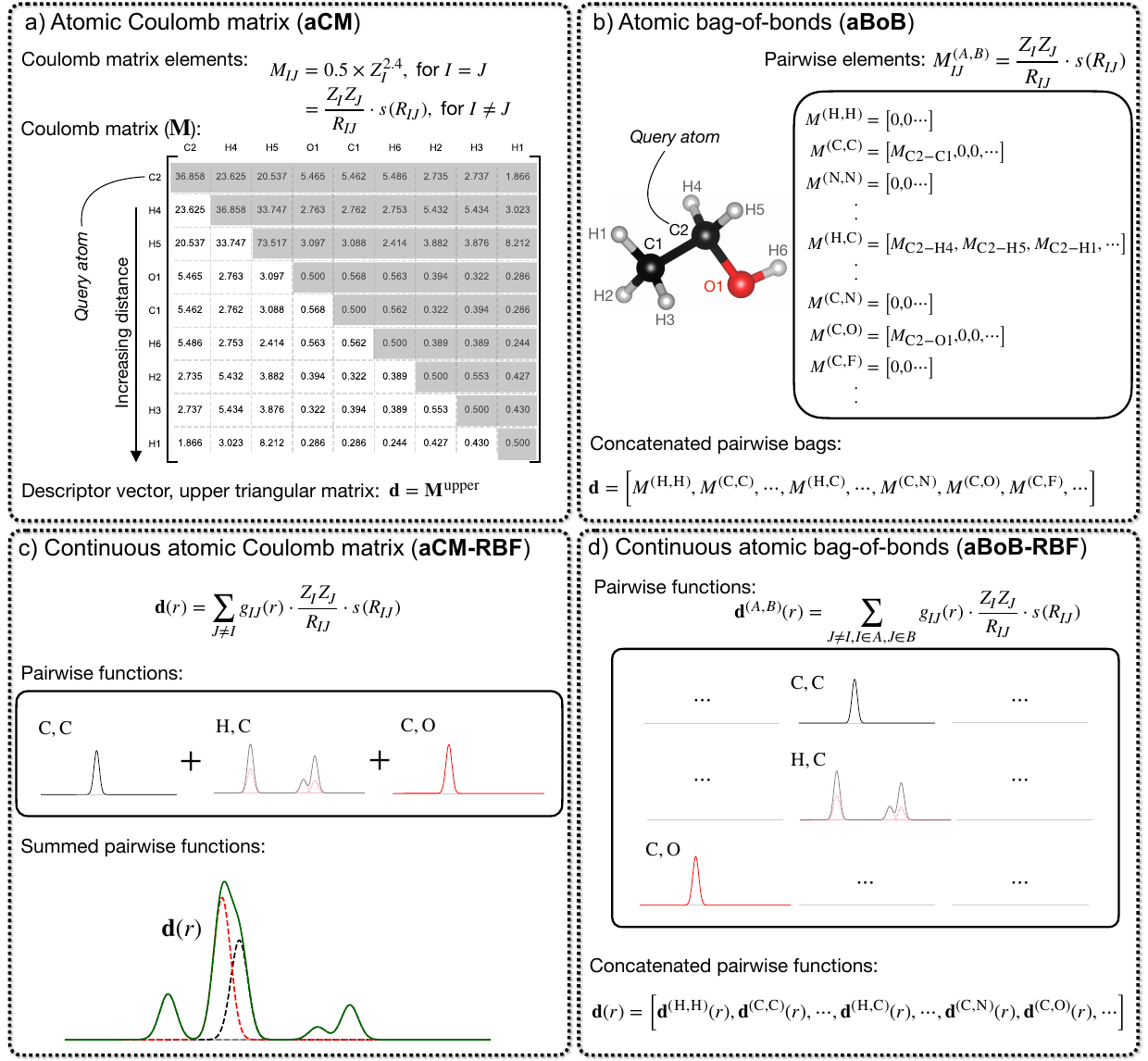}
    \caption{
    Illustration of various descriptors explored in this study.
    (a) aCM: Atomic version of the Coulomb matrix, $M_{IJ}$, where the first index is the query atom, {\it I}, followed by the neighboring atoms sorted in increasing distance. $M_{IJ}$ are calculated using Eq.~\ref{eq:cm_mij} and the upper triangular matrix is vectorized.
    (b) aBoB: aCM elements bagged as pairwise combinations of atom types ($A$ and $B$), $M_{IJ}^{(A,B)}$, where $I \in A $ and $J \in B$ as defined in Eq.~\ref{eq:abob_mij}. Zeros are appended to each bag to ensure constant descriptor sizes across molecules. These bags are then concatenated in a particular sequence to construct the descriptor. 
    (c) aCM-RBF: Continuous version of aCM calculated using Eq.~\ref{eq:con_acm}. For a query atom, $I$, aCM is multiplied by a radial basis function (RBF) 
    and summed over all the neighboring atoms $J$ to obtain the continuous function, ${\bf d}(r)$.
    (d) aBoB-RBF: Continuous version of aBoB calculated using Eq.~\ref{eq:con_abob_mab}, resulting in radial functions for each combination of atom types, ${\bf d}^{(A,B)}(r)$.}
    \label{fig:descriptor_plot}
\end{figure*}

In this study, we investigate the accuracy of descriptors by focusing on two key aspects: the transformation from discrete to continuous representations and the role of many-body effects captured through the nearest-neighbor chemical environment. Continuous descriptors are of constant size~\cite{collins2018constant} and are typically represented on radial or angular grids~\cite{becker1977synthesis} that encode the atomistic surroundings of a query atom. In contrast, discrete descriptors are counted entity-wise (atomwise, bondwise, or via higher-order motifs), making them chemically intuitive but variable in size across molecules. For molecular applications, continuous descriptors offer a physically grounded representation of local environments, while discrete descriptors emphasize interpretable structural features that can be directly compared across systems. Despite their importance, a systematic evaluation of discrete versus continuous formulations in chemical property prediction remains lacking. Beyond chemistry, discrete representations have been widely explored. For example, human language and image processing provide canonical cases where information is encoded as a discrete set of symbols or pixels, respectively~\cite{cartuyvels2021discrete}. Meyer et al. recently showed that large discrete representations can improve performance in continuous reinforcement learning~\cite{meyer2025harnessing}. 
Descriptor performance can also be improved by incorporating higher-order many-body interaction terms, which enhance both accuracy and size efficiency~\cite{huang2020quantummachine}. A well-known example is the 
spectrum of London Axilrod--Teller--Muto 
(SLATM) descriptor~\cite{huang2020quantum}, which includes the three-body Axilrod--Teller--Muto (ATM) potential. Similarly, the bonds and angles ML (BAML) descriptor~\cite{huang2016communication} encodes atom, bond, angle, and torsional terms derived from the many-body expansions of the universal force field (UFF).

We revisit the atomic Coulomb matrix (aCM)~\cite{rupp2012fast}, introduce the atomic variant of bag-of-bonds (BoB)~\cite{hansen2015machine}, denoted aBoB, and develop their continuous counterparts as illustrated in Figure~\ref{fig:descriptor_plot}. Both aCM and aBoB are based on two-body Coulombic interactions, whereas the atomic SLATM (aSLATM)~\cite{huang2020quantum} additionally incorporates three-body Axilrod–Teller–Muto (ATM) terms. We freshly implement aSLATM by integrating the three-body contribution in the $\cos\theta$ domain rather than the $\theta$ domain, leading to a significant reduction in descriptor size. To enhance representability, we further augment each atomic descriptor with contributions from neighboring atomic centers.

\begin{table*}[htbp]
    \centering
    \caption{
    Optimized parameters used in various descriptors are listed, along with their corresponding units. 
    Empty cells indicate that the particular parameter is not available for the corresponding descriptor. 
    Further details on benchmarking are provided in the SI; see 
    Table~S1 ($\Delta t$),
    Table~S2 ($\sigma_{\rm rad}$ and $\sigma_{\rm ang}$),
    Table~S1 ($R_{\rm min}$),
    Table~S3 and Figure~S1 ($R_{\rm max}$),
    Table~S4 and Figure~S2 ($\Delta r$),
    Table~S5 (for $\beta$), and
    Table~S6 ($r_{\rm cut}$).
    } 
    \label{tab:descriptor_para}
    \begin{threeparttable}
        \begin{tabular}{ll c lllll c l}
            \hline
            Parameter & Definition & \text{ } &  \multicolumn{5}{l}{Optimal parameters for various descriptors} & \text{ } & unit\\
            \cline{4-8}
            &  &&  aCM($n$) & aCM-RBF($n$) & aBoB($n$) & aBoB-RBF($n$) & aSLATM($n$) & \\
            \hline
            $\Delta t$            & Angular grid size                                   &&    &      &    &      & 0.05 && none  \\
            $\sigma_{\rm rad}$  & Width of RBF (Eq.~\ref{eq:rbf_2})                   &&    & 0.05 &    & 0.05 & 0.05 && \AA{} \\
            $\sigma_{\rm ang}$  & Width of ABF (Eq.~\ref{eq:abf})                     &&    &      &    &      & 0.05 && none \\
            $R_{\rm min}$         & Minimum of radial domain                            &&    &  0.5 &    &  0.5 &  0.5 && \AA{} \\
            $R_{\rm max}$         & Maximum of radial domain                            &&    &  6.0 &    &  6.0 &  6.0 && \AA{} \\
            $\Delta r$            & Radial grid size                                    &&    & 0.02 &    & 0.05 & 0.05 && \AA{} \\    
            $\beta$               & Scaling function exponent (Eq.~\ref{eq:scaling})    && 3  &    3 & 3  &    3 &    6 && none  \\
            $r_{\rm cut}$         & Radial cutoff  (Eq.~\ref{eq:cos})                   && 2.0&  2.0 & 2.0&  2.0 & 2.0  && \AA{} \\
            \hline
        \end{tabular}
    \end{threeparttable}
\end{table*}

\subsubsection{Atomic Coulomb matrix (aCM)\label{sec:acm}}
The CM is an $N_{\rm atm} \times N_{\rm atm}$ symmetric matrix, where $N_{\rm atm}$ is the number of atoms in a molecule. Its elements are defined as follows: off-diagonal terms are given by Coulombic interactions, while diagonal terms are polynomial fits of the nuclear charge (i.e. atomic number) $Z$ to the free atom’s total energy~\cite{rupp2012fast}:  
\begin{eqnarray}
    M_{IJ} & = & \tfrac{1}{2} Z_{I}^{2.4}, \quad I = J, \nonumber \\
           & = & \frac{Z_{I}Z_{J}}{R_{IJ}}, \quad I \neq J,
   \label{eq:cm}
\end{eqnarray}
where $Z_I$ is the atomic number and $R_{IJ}$ is the interatomic distance between nuclei $I$ and $J$. The CM is invariant to translation, rotation, and reflection, but not to the permutation of atomic indices. To address this, row-sorted, column-sorted, eigenspectrum-based, and randomly permuted CMs have been introduced~\cite{hansen2013assessment}. Barker et al. further localized CM representations for use in Gaussian approximation potentials (GAP), resulting in localized-CM, decaying-CM, and reduced-CM variants~\cite{Barker2017}.  

To represent atomic environments, the aCM can be constructed. Tripathy et al. employed aCM descriptors to predict $1s$ core-electron binding energies of small organic molecules~\cite{tripathy2024chemical}. In aCM, the target atom is placed in the first index, and the remaining atoms are ordered by increasing distance from it. 

To reduce the contribution of distant atoms, a scaling function is introduced:
\begin{eqnarray}
    s(R_{IJ}) = \frac{1}{R_{IJ}^{\beta}},  
    \label{eq:scaling}
\end{eqnarray}
where the optimal value of $\beta$ is reported in Table~\ref{tab:descriptor_para}. The damped CM elements then read:  
\begin{eqnarray}\label{eq:cm_mij}
    M_{IJ} & = & \tfrac{1}{2} Z_{I}^{2.4}, \quad I = J, \nonumber \\
           & = & \frac{Z_{I}Z_{J}}{R_{IJ}} \cdot s(R_{IJ}), \quad I \neq J.
\end{eqnarray}  

To ensure consistent dimensionality across molecules, dummy atoms with $Z=0$ are added so that all aCM descriptors match the maximum number of atoms in the dataset. For QM9, this corresponds to 29 atoms (nonane, C$_9$H$_{20}$). Thus, smaller molecules are zero-padded to maintain uniform descriptor size. An illustration of the aCM construction is provided in Figure~\ref{fig:descriptor_plot}a.

Since the matrix is symmetric, only the upper (or lower) triangular part is vectorized to form the descriptor:  
\begin{equation}
    {\bf d}_I = {\bf M}^{\rm upper}.
\end{equation}  
    
\subsubsection{Atomic bag-of-bonds (aBoB)\label{sec:abob}}
The BoB descriptor shares the same underlying formulation as the CM but differs in representation. Instead of forming a matrix, each pairwise interaction is placed into element-specific bags $(A,B)$, where $A$ and $B$ denote the atom types. 
For consistent dimensionality across the dataset, each bag is zero-padded. The concatenation of all bags in a prescribed sequence yields BoB. Unlike CM, which encodes all pairwise interactions in a matrix, BoB groups them elementwise into fixed bags, allowing the ML model to distinguish atom-type pairs more directly.  

In the localized version, aBoB, only the interactions involving the target atom $I$ and all other atoms in the molecule are included. To account for the reduced influence of distant atoms, a damping factor is applied:  
\begin{equation}\label{eq:abob_mij}
    M^{(A,B)}_{IJ} = \frac{Z_I^A Z_J^B}{R_{IJ}} \cdot s(R_{IJ}),\quad \forall \ I \in A, \ J \in B, \ J \neq I
\end{equation}  
where $s(R_{IJ})$ is the distance-dependent scaling function as defined in Eq.~\ref{eq:scaling}. The resulting pairwise bags are padded with zeros and concatenated in a fixed sequence to form the aBoB descriptor:  
\begin{equation}
    {\bf d}_{I} = [M^{(A,B)}_{IJ}] .
\end{equation}  

For the QM9 dataset, $A,B \in \{\mathrm{H}, \mathrm{C}, \mathrm{N}, \mathrm{O}, \mathrm{F}\}$, giving a total of 15 possible pairwise bags: HH, CC, NN, OO, FF, HC, HO, HN, HF, CO, CN, CF, NO, NF, and OF. Figure~\ref{fig:descriptor_plot}b illustrates the construction of the aBoB descriptor.

\subsubsection{Continuous atomic Coulomb matrix (aCM-RBF)}\label{sec:acm-rbf}

To construct a continuous version of the discrete aCM descriptor, the pairwise elements are replaced with smooth functions of the interatomic distance. Each contribution combines a term from aCM, with a normalized Gaussian radial basis function (RBF) centered at $R_{IJ}$ and a distance-dependent scaling factor:  
\begin{equation}\label{eq:rbf_2}
        g_{IJ}(r) = \frac{1}{\sigma_{\rm rad} \sqrt{2\pi}} 
        \exp\!\left[-\frac{(r - R_{IJ})^2}{2\sigma_{\rm rad}^2}\right],
\end{equation}
where $\sigma_{\rm rad}$ controls the Gaussian width and thus the spread of the function given in Table~\ref{tab:descriptor_para}.

RBFs such as Gaussians are widely used in molecular descriptors, for example, in SLATM~\cite{huang2020quantum} and in Fourier-series expansions of atomic radial distribution functions~\cite{von2015fourier}. The resulting continuous representation, denoted aCM-RBF, is given by  
\begin{equation}\label{eq:con_acm}
     {\bf d}_{I}(r) = \sum_{J \neq I} g_{IJ}(r) \cdot M_{IJ},
\end{equation}
where $M_{IJ}$ is defined in Eq.~\ref{eq:cm_mij}. A schematic illustration of the aCM-RBF construction is provided in Figure~\ref{fig:descriptor_plot}c.

\subsubsection{Continuous atomic bag-of-bonds (aBoB-RBF)\label{sec:abob-rbf} } 

For the continuous version of aBoB, aBoB-RBF, the same procedure as continuous aCM-RBF is followed. First, pairwise elements, $M_{IJ}^{(A,B)}$ are calculated between the target atom index $I$ (with $A$-type atom) and the rest of the atoms (with indices $J$, $B$-type atom) and then subjected to distribute around the normalized Gaussian function, $g_{IJ}(r)$ and a scaling function, $s(R_{IJ})$. For each type of pairwise function, all the individual functions are summed. Which will provide the pairwise function given as 
\begin{equation}\label{eq:con_abob_mab}
        {\bf d}^{(A,B)}(r) = \sum_{I \in A, J \in B, J \neq I} g_{IJ}(r) \cdot M^{(A,B)}_{IJ},
\end{equation}
where $M^{(A,B)}_{IJ}$ is defined in Eq.~\ref{eq:abob_mij}.
Then these pairwise functions are concatenated in a particular sequence to construct the aBoB-RBF descriptor.
\begin{equation}
        {\bf d}_{I}(r) = [{\bf d}^{(A,B)}(r) ] \text{ } \forall \ I \in A, \ J \in B, \ J \neq I.
\end{equation}
The aBoB-RBF descriptor will be the number of bags (i.e. number of unique $(A,B)$ pairs) times larger than aCM-RBF. 
A pictorial description to construct the aBoB-RBF descriptor is shown in Figure~\ref{fig:descriptor_plot}d.
The formulas for the lengths of the aCM, aBoB, aCM-RBF, and aBoB-RBF descriptors are summarized in Table~S7 of the SI.

\subsubsection{Physical motivation for multiplying the Coulomb matrix with a scaling function}
The scaling factor $s(R_{IJ}) = 1/R_{IJ}^{\beta}$ (see Eq.~\ref{eq:scaling}) is not merely a numerical device for suppressing distant atomic contributions. 
It has a direct physical motivation in the long-range behavior of NMR shielding, $\sigma(R)$.  
The NMR shielding tensor is a second-order magnetic response property that couples the external magnetic field to induced electronic currents.  
Analyses of atom--atom interactions by Barszczewicz et al.~\cite{barszczewicz1996long} showed that the 
deviation of the shielding of an interacting atom, from the free-atom value, decays asymptotically as $1/ R^n$,
with $n = 3, 4,$ and $6$ for the anisotropy of the shielding tensor (dipolar contribution), atom--ion interactions (induction-dominated), and the isotropic shielding of two neutral atoms (dispersion-dominated), respectively.  
These inverse-power laws reflect the fact that shielding is controlled not by static electrostatics but by induced magnetic fields and current-density response, which decay much more rapidly with distance.  
Thus, a descriptor whose radial decay lies within the physically relevant range $1/R^3$--$1/R^6$ is far more consistent with magnetic response theory than an undamped Coulomb matrix suitable to map to molecular total energy.

Multiplying the Coulomb interaction $Z_I Z_J / R_{IJ}$ by an additional factor of $1/R_{IJ}^3$ yields descriptor elements proportional to placing the effective decay rate between the dominant long-range regimes.
Importantly, for typical organic molecules such as those in QM9, the electronic environments around heteroatoms and substituted carbons resemble quasi-ion or quasi-dipole centers rather than spherically symmetric neutral atoms.  
Local bond polarity, electronegativity differences, and hybridization generate partial charges and permanent or induced dipoles.  
In such situations, the induction mechanism, which leads to an $1/R^4$ scaling of the isotropic shielding, is expected to dominate over the dispersion-driven $1/R^6$ decay characteristic of strictly neutral closed-shell atoms.  
Hence, the $1/R^4$ behavior implicit in the scaled Coulomb matrix is not only physically justified but closely aligned with the shielding response expected for realistic molecular environments.

\subsubsection{Atomic Spectrum of London Axilrod--Teller--Muto (aSLATM)}\label{sec:aslatm}

The SLATM descriptor, and its atomic version aSLATM, encode molecular environments using one-, two-, and three-body terms~\cite{huang2020quantum}. For an atom, $I$, the one-body contribution is the atomic number:
\begin{equation}\label{eq:aslatm_1body}
        {\bf d}_I^{(A)} = Z_{I}.
\end{equation}

The two-body (radial) term is given by
\begin{equation}\label{eq:aslatm_2body}
        {\bf d}_I^{(A,B)}(r) = \frac{1}{2} Z_I \sum_{J \ne I} Z_J \cdot g_{IJ}(r) \cdot s(r),
\end{equation}
where $g_{IJ}(r)$ is the normalized Gaussian RBF (Eq.~\ref{eq:rbf_2}) and $s(r)=1/r^{\beta}$ is the distance-dependent scaling. For the radial part, $\beta=6$ is used, consistent with the leading-order dissociative tail of the London potential.  

The three-body (angular) term accounts for ATM contributions:  
\begin{equation}\label{eq:aslatm_3body}
        {\bf d}_I^{(A,B,C)}(\theta) = \frac{1}{3} Z_I \sum_{K \ne J \ne I} Z_J Z_K \cdot g_{IJK}(\theta) \cdot h(\theta, {\bf R}_{IJ}, {\bf R}_{IK}),
\end{equation}
where $g_{IJK}(\theta)$ is the angular basis function (ABF) and $h(\theta, {\bf R}_{IJ}, {\bf R}_{IK})$ is the ATM potential:
\begin{equation}\label{eq:atm}
        h(\theta, {\bf R}_{IJ}, {\bf R}_{IK}) = 
        \frac{1 + \cos  \theta \cdot \cos \theta_{JKI} \cdot \cos \theta_{KIJ}}
             {(R_{IJ} R_{JK} R_{KI})^3}.
\end{equation}

Thus, the full aSLATM descriptor is formed by concatenating the one-body, two-body, and three-body terms:  
\begin{equation}
      {\bf d}_{I}(r,\theta) = \big[{\bf d}_I^{(A)}, {\bf d}_I^{(A,B)}(r), {\bf d}_I^{(A,B,C)}(\theta)\big].
\end{equation}
For the QM9 dataset, $A,B,C \in \{\mathrm{H}, \mathrm{C}, \mathrm{N}, \mathrm{O}, \mathrm{F}\}$, giving 5 unique atom types, 15 unique $(A,B)$ pairs, and 35 unique $(A,B,C)$ triplets. A common implementation of SLATM and aSLATM is available in the \texttt{QML} package~\cite{QML}.  

In this study, we introduce a modification to the angular term by integrating over the $\cos \theta$ domain rather than $\theta$:  
\begin{equation}\label{eq:aslatm_t}
        {\bf d}_I^{(A,B,C)}(t) = \frac{1}{3} Z_I \sum_{J \ne K \ne I} Z_J Z_K \cdot g_{IJK}(t) \cdot h(t, {\bf R}_{IJ}, {\bf R}_{IK}),
\end{equation}
with $t \equiv \cos \theta$. Representing the angular term in $\cos \theta$ reduces the integration domain to $[-1,1]$, making it compatible with standard quadrature schemes for trigonometric functions and significantly reducing descriptor size. The ATM potential and angular RBFs are then expressed as  
\begin{equation}\label{eq:atm_t}
        h(t, {\bf R}_{IJ}, {\bf R}_{IK}) = \frac{1 + t \cdot t_{JKI} \cdot t_{KIJ}}{(R_{IJ} R_{JK} R_{KI})^3},
\end{equation}
\begin{equation}\label{eq:abf}
        g_{IJK}(t) = \frac{1}{\sigma_{\rm ang} \sqrt{2\pi}} 
        \exp \!\left[ -\frac{(t - t_{IJK})^2}{2\sigma_{\rm ang}^2}\right].
\end{equation}

 The final descriptor is expressed as  
\begin{equation}
      {\bf d}_{I}(r,t) = \big[{\bf d}_I^{(A)}, {\bf d}_I^{(A,B)}(r), {\bf d}_I^{(A,B,C)}(t)\big].
\end{equation}

All radial and angular basis widths, step sizes for integration, and domain limits are summarized in Table~\ref{tab:descriptor_para}.

\subsubsection{Neighbor atom information encoded descriptors: 
aCM($n$), aBoB($n$), aCM-RBF($n$), aBoB-RBF($n$), and aSLATM($n$)
\label{sec:neighbors_information}}
To more accurately capture the chemical environment of an atom, we extend the parent descriptors by incorporating information from the $n$ nearest neighbors. The resulting descriptor is denoted ${\bf d}(n)$, where ${\bf d}(0)$ corresponds to the parent descriptor of the query atom alone, and ${\bf d}(n>0)$ additionally includes contributions from the $n$ closest atoms, ordered by increasing distance. If fewer than $n$ neighbors exist, zero values are appended to preserve dimensionality. While this approach increases descriptor size, it provides a richer representation of the local chemical environment. 
This scheme allows systematic inclusion of neighboring environments, which becomes especially important for complex systems such as solvated systems, organometallics, or transition-metal complexes. 
The lengths of the aCM($n$), aBoB($n$), aCM-RBF($n$), and aBoB-RBF($n$) descriptors scale linearly with $n$, i.e., they are $n+1$ times the size of their $n=0$ variants listed in Table~S7 of the SI.

\begin{figure}[!htbp]
    \centering
    \includegraphics[width=\linewidth]{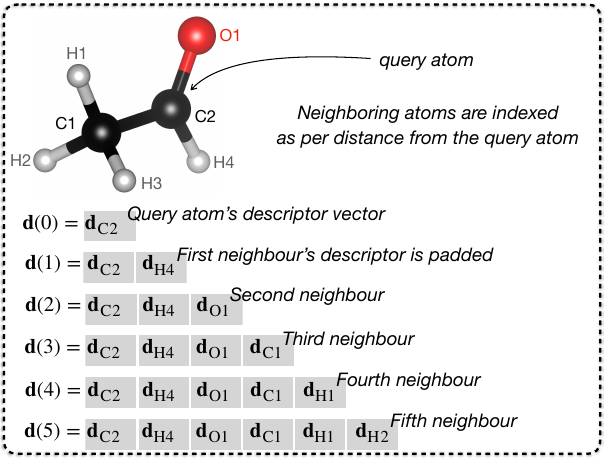}
    \caption{
    Encoding of neighborhood information in atomic descriptors illustrated for the carbonyl carbon (C2) of acetaldehyde. The query atom's descriptor is ${\bf d}(0)$, which is concatenated with the descriptors of the neighboring atoms after damping with the 
    $f_{\rm cos}(d_k; 2.0)$
    function. 
    }
    \label{fig:neighbors}
\end{figure}

As an illustration, consider the C2 atom in acetaldehyde (Figure~\ref{fig:neighbors}). With $n=0$, the descriptor is simply ${\bf d}(0)={\bf d}_{\rm C2}$. For $n=1$, the nearest atom H4 is included: ${\bf d}(1)=\left[ {\bf d}_{\rm C2}, {\bf d}_{\rm H4} \right]$. For $n=2$, both H4 and O1 are added: ${\bf d}(2)=\left[ {\bf d}_{\rm C2}, {\bf d}_{\rm H4}, {\bf d}_{\rm O1}\right]$, and so forth.

Since the influence of neighbors decays with distance, damping functions are applied to weight their contributions. We consider three forms of cutoff functions, namely, exponential, polynomial, and cosine cutoffs:
\begin{equation}\label{eq:exp} 
    f_{\rm exp.}(d_k;\lambda) = e^{-\lambda d_k}, 
    \end{equation} 
\begin{equation}\label{eq:pol} 
    f_{\rm poly.}(d_k;p) = \frac{1}{(1 + d_k)^p}, 
\end{equation} 
\begin{equation}\label{eq:cos} 
    f_{\rm cos.}(d_k; r_{\rm cut}) = \frac{1}{2} \left[ 1 + \cos \left( \pi \frac{d_k}{r_{\rm cut}} \right) \right], 
\end{equation}
where $d_k$ is the distance between the query atom and its $k^{\text{th}}$ neighbor, and $\lambda$, $p$, and $r_{\rm cut}$ are parameters controlling the decay of the respective damping functions.  
The performance of different damping schemes was benchmarked on the 5k subset of QM9, and for the cosine damping function with
$r_{\rm cut}=2.0$~\AA{}, low errors were obtained consistently across all various descriptors as shown in Table~S6.
Hence, this 
$f_{\rm cos}(d_k; 2.0)$
scheme is adopted as the default damping function for all descriptors in this study.
The implementation of the damped neighbour descriptors is described below. For example, in the C2 atom in acetaldehyde as shown in Figure~\ref{fig:neighbors}, with $n=0$, the descriptor is ${\bf d}(0)=\left[ {\bf d}_{\rm C2} \right]$. For $n=1$, with inclusion of the damping function, the descriptor becomes ${\bf d}(1)=\left[ {\bf d}_{\rm C2},\, {\bf d}_{\rm H4}\cdot f_{\rm cos}(R_{\rm C2-H4};2.0) \right]$. Similarly, for $n=2$, we arrive at ${\bf d}(2)=\left[ {\bf d}_{\rm C2},\, {\bf d}_{\rm H4}\cdot f_{\rm cos}(R_{\rm C2-H4};2.0),\, {\bf d}_{\rm O1}\cdot f_{\rm cos}(R_{\rm C2-O1};2.0)\right]$, and so forth. Here, $R_{\rm C2-H4}$ and $R_{\rm C2-O1}$ denote the distance between the query C2-atom and its first and second neighbor atoms, respectively (see Figure~\ref{fig:neighbors}).

In summary, for both the query atom and its neighbors, the scaling function $s$ reduces the weight of descriptor elements as the interatomic distance increases, as illustrated in Fig.~\ref{fig:descriptor_plot}. The damping function $f$ further suppresses the contribution of the entire neighbor descriptor vector for the query atom with increasing distance (Fig.~\ref{fig:neighbors}). Together, $s$ and $f$ ensure that the descriptor of a query atom remains ``size-consistent'', meaning that its representation of the local atomic environment does not change when the overall molecular size increases. This property enables consistent training and prediction across molecules of varying sizes.

%

\subsection{Kernel-Ridge Regression\label{sec:krr}}
Kernel-ridge regression is a regularized least squares regression method~\cite{kung2014kernel}, and one of the most accurate ML frameworks for modeling chemical properties~\cite{rupp2012fast}. 
It has been successfully applied to global molecular properties such as atomization energies~\cite{rupp2012fast,hansen2015machine}, as well as local properties including NMR shielding~\cite{gupta2021revving} and core electron binding energies~\cite{tripathy2024chemical,porcelli2025photoemission}. 

For a query system $q$, the target property $p_q$ is expressed as a linear combination of kernel elements constructed from the descriptor differences between the query and all training set entries:
\begin{equation}\label{eq:krr}
p_q = \sum_{t=1}^{N} c_{t} K_{qt}.
\end{equation}
In this study, the target property is the $^{13}$C isotropic shielding, $\sigma_{\rm iso}(^{13}{\rm C})$. Predicted shielding values are subsequently converted to chemical shifts, $\delta(^{13}{\rm C})$, using Eq.~\ref{eq:chemicalshift}.
The regression coefficients $c_t$ are obtained by solving the linear system
\begin{equation}\label{eq:krr_closed_form_eqn}
\left[{\bf K} + \lambda {\bf I}\right] {\bf c} = {\bf p},
\end{equation}
where ${\bf K}$ is the $N \times N$ kernel matrix, ${\bf I}$ the identity matrix, $\lambda$ the regularization parameter, and ${\bf p}$ the vector of target properties for the $N$ training entries.

The kernel matrix elements are constructed by applying a kernel function to the descriptor difference, $K_{ij} = k(D_{ij})$. To characterize different chemical environments, descriptor differences can be computed using either the $L^1$ (Taxicab or Manhattan) or $L^2$ (Euclidean) norm:
\begin{equation}\label{eq:dij_l1}
||D_{ij}||_1 = ||{\bf d}_i - {\bf d}_j||_1 = \sum_{k=1}^{l} |({\bf d}_i)_k - ({\bf d}_j)_k|,
\end{equation}
\begin{equation}\label{eq:dij_l2}
||D_{ij}||_2 = ||{\bf d}_i - {\bf d}_j||_2 = \sqrt{\sum_{k=1}^{l} \left[ ({\bf d}_i)_k - ({\bf d}_j)_k\right]^2},
\end{equation}
where $l$ is the size of descriptor vectors. 
For continuous descriptors, the descriptor difference can be calculated in the continuous domain ($r \in [R_{\rm min}, R_{\rm max}]$) as a difference of the smooth functions as:
\begin{equation}\label{eq:dij_l1}
||D_{ij}||_1 = \int_{R_{\rm min}}^{R_{\rm max}} {\rm d}r\left| {\bf d}_i(r) - {\bf d}_j (r)\right| ,
\end{equation}
\begin{equation}\label{eq:dij_l2}
||D_{ij}||_2 = \sqrt{ \int_{R_{\rm min}}^{R_{\rm max}} {\rm d}r\left| {\bf d}_i(r) - {\bf d}_j (r)\right|^2}.
\end{equation}
A similar integration can be carried out in the angular domain by transforming to $t=\cos \theta$, with $t \in [-1,1]$, ensuring consistency with the formulation of continuous angular descriptors such as aSLATM.
In practice, both $r$ and $t$ integrations are performed numerically by discretizing the domains into finite steps of $\Delta r$ and $\Delta t$, respectively. This yields discrete sums that approximate the continuous integrals while retaining the smoothness of the descriptor functions.  See Table~\ref{tab:descriptor_para} for optimal values of the parameters involved.

Using these distance measures, the kernel elements are then defined by either a Laplacian or a Gaussian kernel:
\begin{equation}\label{eq:kij}
K_{ij}^{\rm Laplacian} = \exp \left( -\frac{||D_{ij}||_1}{\sigma} \right); \,
K_{ij}^{\rm Gaussian} = \exp \left( -\frac{||D_{ij}||_2^2}{2\sigma^2} \right),
\end{equation}
where $\sigma$ denotes the kernel width hyperparameter. 
In earlier studies on molecular global properties, the Laplacian kernel was found to perform well with discrete descriptors~\cite{hansen2015machine}. More recently, Shiota et al. reported that the relative performance of Laplacian and Gaussian kernels in KRR depends on dataset sampling and training set size~\cite{shiota2024universal}. To assess this in the present context, we evaluated both kernels across all descriptor types for predicting $^{13}$C chemical shifts (see Table~S8 in the SI). In every case, the Laplacian kernel yielded lower prediction errors, which we use in all KKR models reported in this study.

In the KRR framework, two key hyperparameters that control the accuracy of the model are the regularization strength $\lambda$ (Eq.~\ref{eq:krr_closed_form_eqn}) and the kernel width $\sigma$ (Eq.~\ref{eq:kij}). The regularization parameter $\lambda$ ensures the non-singularity of the kernel matrix and penalizes regression coefficients of large magnitudes to mitigate overfitting. Previous studies recommend small values of $\lambda$ to mitigate linear dependency issues for large training set sizes~\cite{ramakrishnan2015many,tripathy2024chemical,gupta2021revving}. In this study, we set $\lambda=10^{-5}$.
The kernel width $\sigma$ controls the spread of similarity across the training set. 
A heuristic strategy for optimizing $\sigma$ sets the minimum kernel value to $K_{ij}^{\rm min}=0.5$ over the kernel range $K_{ij}\in[0,1]$, yielding~\cite{ramakrishnan2015many}:
\begin{equation}\label{eq:sigma_opt_max}
\sigma_{\rm opt}^{\rm Laplacian} = \frac{D_{ij}^{\rm max}}{\log 2},
\qquad
\sigma_{\rm opt}^{\rm Gaussian} = \frac{D_{ij}^{\rm max}}{\sqrt{2 \log 2}}.
\end{equation}

In the QM9NMR study~\cite{gupta2021revving}, improved accuracy was obtained by replacing $D_{ij}^{\rm max}$ with the median descriptor difference $D_{ij}^{\rm median}$. This approach was later generalized for core-electron binding energy modeling~\cite{tripathy2024chemical} by scanning $K_{ij}^{\rm min}$ over a continuous range $\tau \in (0,1)$, leading to the definition of a ``$\tau$-plot'' for kernel width optimization:
\begin{equation}\label{eq:tau_opt}
\sigma_{\rm opt}^{\rm Laplacian} = \frac{D_{ij}^{\rm median}}{\log(1/\tau)},
\qquad
\sigma_{\rm opt}^{\rm Gaussian} = \frac{D_{ij}^{\rm median}}{\sqrt{2 \log(1/\tau)}}.
\end{equation}
In this study, we calculated $D_{ij}^{\rm median}$ using the training entries. This quantity can also be determined using a fixed, reasonably large training set (say, of size 1k) and applied to larger sets. 
As shown in Figure~S3 of the SI, out-of-sample errors are minimized for $\tau \approx 0.3-0.5$ for more accurate descriptors such as  
aBoB-RBF(4) and aSLATM(0). Accordingly, we adopt $\tau=0.5$ in this study, as this value is consistent with the previous KRR-modeling of NMR shielding~\cite{gupta2021revving}.

\section{Results and Discussions}\label{sec:results}

We evaluate the performance of the proposed descriptors: aCM($n$), aBoB($n$), aCM-RBF($n$), and aBoB-RBF($n$), for predicting $\sigma(^{13}{\rm C})$, comparing them with aSLATM($n$) and previously reported results. Initial assessments of various parameters involved in the ML models were carried out on the 5k subset of QM9 (see Figure~\ref{fig:qm9_13c_dist}), comprising $\sim$32k C atoms, with 8k used for training and 2k for testing 
(see Table~\ref{tab:descriptor_para}). 

For benchmarking against prior studies, we employ a random sample of 150k carbon atoms from the full QM9NMR dataset ($\sim$832k C atoms in total), split into 100k training and 50k test entries. Furthermore, external validation is performed by applying QM9NMR-trained models to the Drug12 and Drug40 datasets, followed by evaluation on 208 pyrimidinone derivatives and 200 GDBm molecules. 

As described in Section~\ref{sec:dataset}, target NMR shieldings were calculated at the mPW1PW91/6-311+G(2\textit{d,p}) level using B3LYP/6-31G(2\textit{df,p}) geometries, whereas the KRR models require only PM7-optimized minimum energy geometries as input.

\begin{figure}[!htbp]
    \centering
    \includegraphics[width=\linewidth]{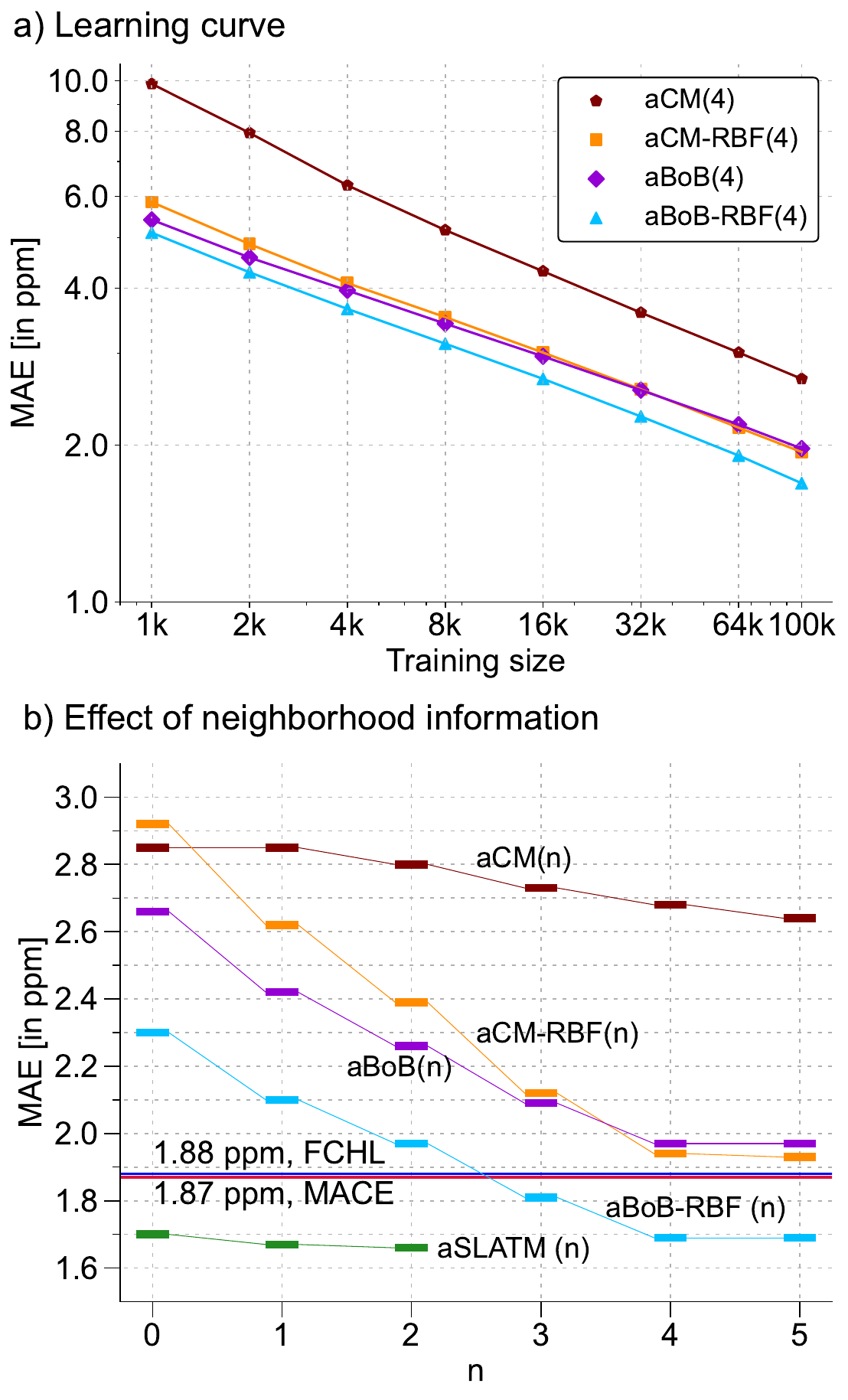}
    \caption{a) Learning curves on the log-log scale showing mean absolute errors (MAEs in ppm) for predicting $^{13}$C-NMR $\sigma_{\rm iso}$ values with increasing training set sizes. MAEs are calculated for 50k out-of-sample atoms in the QM9NMR dataset. All descriptors have information of the four nearest neighbor atoms (i.e., ${\bf d}(4)$). 
    b) MAEs for a train:test split of 100k:50k for various descriptors with information of $n$ nearest atoms encoded. Previously reported values of 1.88 (FCHL, \RRef{gupta2021revving}) and 1.87 (MACE-OFF23-small, \RRef{shiota2024universal}) are shown as blue and red horizontal lines. Values for the descriptors studied in this work are shown for various $n$: aCM (maroon), aBoB (violet), aCM-RBF (orange), aBoB-RBF (skyblue), and aSLATM (green).}
    \label{fig:lc_benchmark}
\end{figure}

    \begin{figure*}[!htbp]
        \centering
        \includegraphics[width=\linewidth]{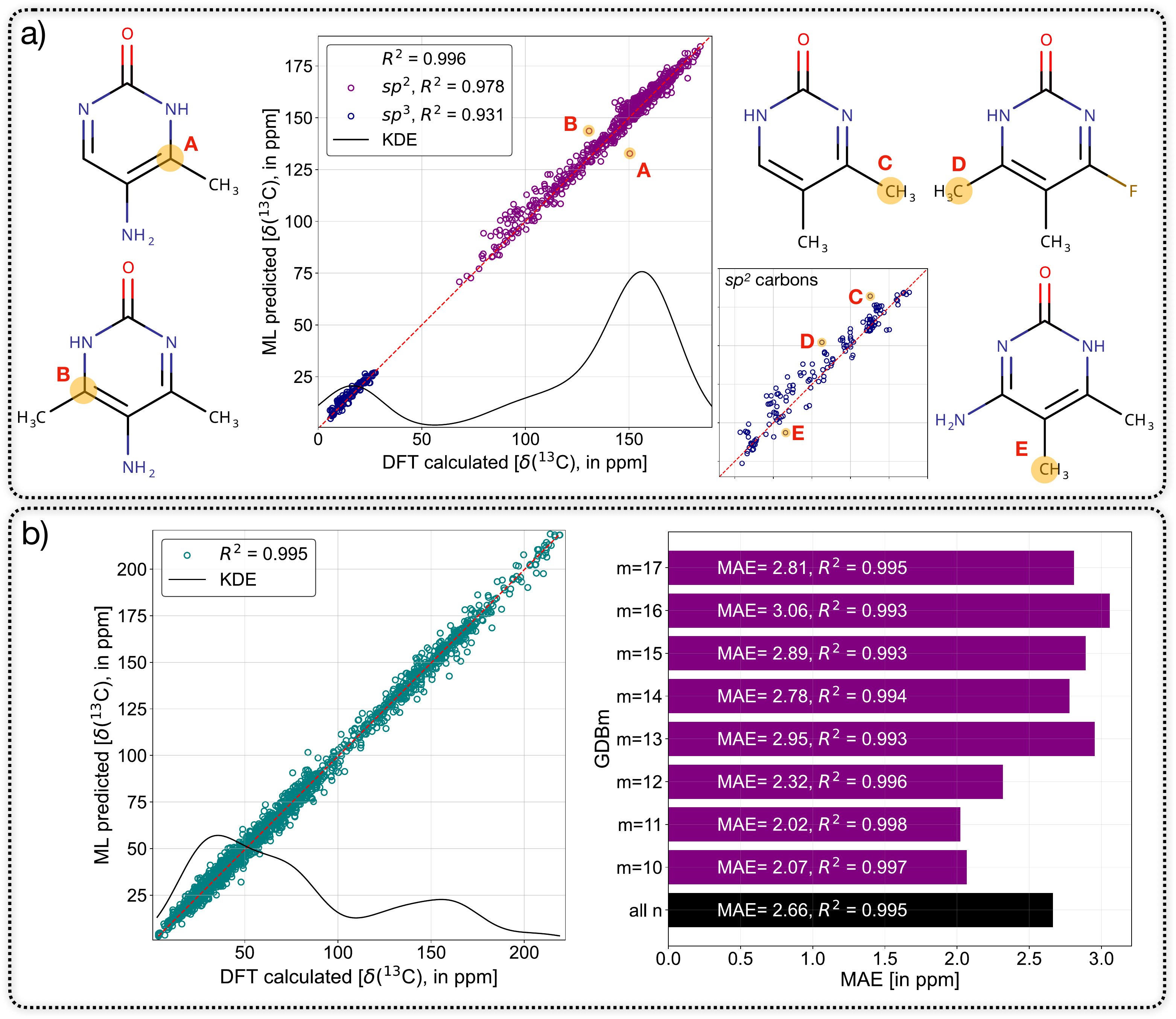}
        \caption{Analysis of errors of ML models predicting $^{13}C$ chemical shifts, $\delta(^{13}{\rm C})$, for pyrimidinone and GDBm validation sets. 
        a) Scatterplot of ML-predicted vs. DFT-calculated $\delta(^{13}{\rm C})$ values for 208 pyrimidinone molecules. Results are shown for  {\rm sp$^3$} and {\rm sp$^2$} C atoms. 
        The density plot shows the distribution of the DFT results. 
        Extreme values with larger MAEs are highlighted as A--E, and these atoms are shown in the corresponding molecules. 
        b) Scatterplot of ML-predicted vs. DFT-calculated $\delta(^{13}{\rm C})$ values for 200 GDBm molecules along with the distribution of DFT values. MAEs (in ppm) for each value of $m$ are shown along with the coefficient of determination, $R^2$, as a bar plot.}
        \label{fig:pyrimidinone_GDB_analysis}
    \end{figure*}

\subsection{Prediction accuracy on QM9 molecules}

To assess how well the ML models capture the chemical environment with increasing data, we plot $\log$–$\log$ learning curves of mean absolute error (MAE) versus training set size in Figure~\ref{fig:lc_benchmark}a. The monotonically decreasing MAEs for aCM(4), aCM-RBF(4), aBoB(4), and aBoB-RBF(4) confirm that local atomic representations effectively encode the chemical environment and improve predictions of $\delta(^{13}{\rm C})$ as training data grows. Notably, the transition from the discrete aCM(4) to the continuous aCM-RBF(4) yields a pronounced reduction in error, whereas for aBoB(4) versus aBoB-RBF(4), the improvement is smaller but still noticeable, indicating that continuous representations capture shielding information more efficiently.
To further quantify learning efficiency, we fit the learning curves to the power-law form
$\log(\mathrm{Error}) = a + b,\log(N)$.
The fitted slopes for aCM(4), aBoB(4), aCM-RBF(4), and aBoB-RBF(4) are $-0.281$, $-0.216$, $-0.236$, and $-0.260$, respectively, indicating that all four descriptors exhibit comparable learning behavior. In practice, the slope of a learning curve is strongly problem dependent and reflects the underlying complexity of the target function. For molecular machine-learning tasks such as NMR shielding, where the property depends on detailed, high-dimensional local environments, learning curves typically decay more slowly. The slopes observed here (approximately $-0.25$, corresponding to $\mathrm{Error}\propto 1/N^{0.25}$) are therefore consistent with expectations for this type of problem, rather than faster decays such as $\mathrm{Error}\propto 1/N$ or $\mathrm{Error}\propto 1/\sqrt{N}$. The strictly monotonic decrease in MAE confirms stable and systematic learning as the dataset grows across all descriptors.
    
The impact of incorporating neighbor information is shown in Figure~\ref{fig:lc_benchmark}b using a 100k:50k train/test split. For all four descriptors (aCM, aCM-RBF, aBoB, and aBoB-RBF), the MAE decreases steadily as neighbors are added from $n=0$ to $n=4$, but begins to saturate at $n=5$. By contrast, the aSLATM($n$) descriptor shows little benefit from additional neighbors. This can be attributed to the fact that aSLATM($0$) already incorporates explicit three-body terms, such that neighbor inclusion does not provide additional unique information. This suggests a close connection between many-body terms, such as the three-body ATM term in SLATM, and atomic neighborhood effects. 

        \begin{table*}[htbp]
            \centering
            \caption{MAEs of ML models with different descriptors trained on the chemical shift of 100k C atoms in QM9NMR. 
            MSE, MAE, and SDE are mean signed error, mean absolute error, and standard deviation of the error, respectively. Error metrics from previous studies are provided for comparison. 
            All values are in ppm.}
            \label{tab:test_validation}
            \begin{tabular}{l c lll c lll c lll c l}
            \hline
            Descriptor              & & \multicolumn{4}{l}{QM9NMR (50k)} & \multicolumn{4}{l}{Drug12 (182)} & \multicolumn{4}{l}{Drug40 (381)} & Reference \\
            \cline{3-5}
            \cline{7-9}
            \cline{11-13}
                                && MSE     & MAE  & SDE  && MSE     & MAE  & SDE  && MSE   & MAE  & SDE    &&  \\
            \hline
            FCHL                &&         & 1.88 &      &&         & 4.2  &      &&         & 3.7  &      && \RRef{gupta2021revving} \\  
            MACE-OFF23-small    &&         & 1.87 &      &&         & 3.85 &      &&         & 2.83 &      && \RRef{shiota2024universal} \\
            aBoB-RBF(3)         && $+$0.01 & 1.81 & 2.78 && $-$1.27 & 3.71 & 4.84 && $-$0.50 & 2.84 & 4.04 && this study \\
            aBoB-RBF(4)         && $+$0.01 & 1.69 & 2.65 && $-$1.03 & 3.43 & 4.58 && $-$0.49 & 2.80 & 4.01 && this study \\
            aSLATM(0)           && $-$0.02 & 1.70 & 2.68 && $-$1.25 & 3.51 & 4.73 && $-$0.39 & 2.77 & 4.35 && this study \\
            aSLATM(1)           && $+$0.01 & 1.67 & 2.62 && $-$0.91 & 3.79 & 5.26 && $-$0.35 & 2.85 & 4.48 && this study \\
            aSLATM(2)           && $+$0.00 & 1.66 & 2.61 && $-$0.61 & 3.73 & 5.19 && $-$0.19 & 2.80 & 4.32 && this study \\
            \hline
            \end{tabular}
        \end{table*}

A consistent decrease in MAE is observed in Figure~\ref{fig:lc_benchmark}b when moving from discrete vector descriptors to their continuous function forms, as seen in the comparisons of aCM($n$) vs. aCM-RBF($n$) and aBoB($n$) vs. aBoB-RBF($n$). In nearly all cases, the continuous versions outperform their discrete counterparts, with the only exception being aCM(0) versus aCM-RBF(0). Improvements are more pronounced for aCM-RBF($n$) than for aCM($n$), while the gains for aBoB-RBF($n$) relative to aBoB($n$) are nearly constant for each $n$. These results demonstrate that both neighbor inclusion and continuous functional representations enhance descriptor accuracy.
 
As shown in Figure~\ref{fig:lc_benchmark}b, aBoB-RBF(4) achieves lower MAEs than previously benchmarked models (see Table~\ref{tab:test_validation}). Notably, aBoB-RBF reaches the accuracy of FCHL and MACE with $n=3$, and with $n=4$ it matches the performance of aSLATM(0). The comparable errors of aSLATM(0) and aBoB(4) indicate that SLATM’s explicit three-body terms effectively capture nearest-neighbor effects, explaining why further neighbor inclusion in aSLATM does not improve accuracy.

We also modeled $^{13}$C chemical shifts from the QM9NMR dataset using the $\Delta$-ML approach~\cite{ramakrishnan2015big}. When using HF/STO-3G baseline  computed on PM7 geometries and the aBoB-RBF(4) descriptor, the MAEs on a 50k hold-out test set decrease to 1.66, 1.46, 1.29, and 1.17 ppm for training set sizes of 16k, 32k, 64k, and 100k, respectively. Using mPW1PW91/STO-3G baseline calculations on the same PM7 geometries yields very similar performance, with MAEs of 1.66, 1.46, 1.28, and 1.16 ppm for the corresponding training sizes.
For comparison, Gupta et al. reported an MAE of 1.36 ppm for gas-phase $^{13}$C NMR chemical shielding using the FCHL descriptor in a 100k $\Delta$-ML model~\cite{gupta2021revving}. Notably, with the aBoB-RBF(4) descriptor, the MAE already reaches 1.46 ppm using only 32k training entries, and approaches the 1 ppm accuracy threshold required for structure elucidation when trained on a 100k set. These results demonstrate the strong applicability of the continuous aBoB-RBF(4) descriptor for $\Delta$-ML modeling of NMR properties.

The aBoB-RBF($n$) descriptor was further evaluated for modeling $^{15}$N NMR chemical shifts for the nitrogen-containing molecules in the QM9NMR dataset. As $n$ increases from 0 to 5, the MAEs decrease to 4.08, 3.44, 2.99, 2.78, 2.73, and 2.75 ppm, respectively, using a training set of 100k molecules and an out-of-sample test set of 30k molecules. Using the same number of data points, Shiota et al. report an MAE of 3.42 ppm with the M3GNet GNN-TL embedding descriptor~\cite{shiota2024universal}. The systematic decrease in MAE with increasing $n$ indicates that the aBoB-RBF descriptors effectively capture the local atomic and electrostatic environments of nitrogen atoms in these molecules. Notably, the MAEs for $^{15}$N converge at $n=3$, whereas for $^{13}$C they converge at $n=4$, which can be rationalized in terms of differences in typical coordination numbers or valencies of the two atom types.

Overall, aBoB-RBF(4), with an MAE of 1.69 ppm, surpasses the previously reported benchmarks of 1.88 ppm for FCHL~\cite{gupta2021revving} and 1.87 ppm for the MACE-OFF23-small descriptor~\cite{shiota2024universal}.

\begin{figure}[!htbp]
    \centering
    \includegraphics[width=\linewidth]{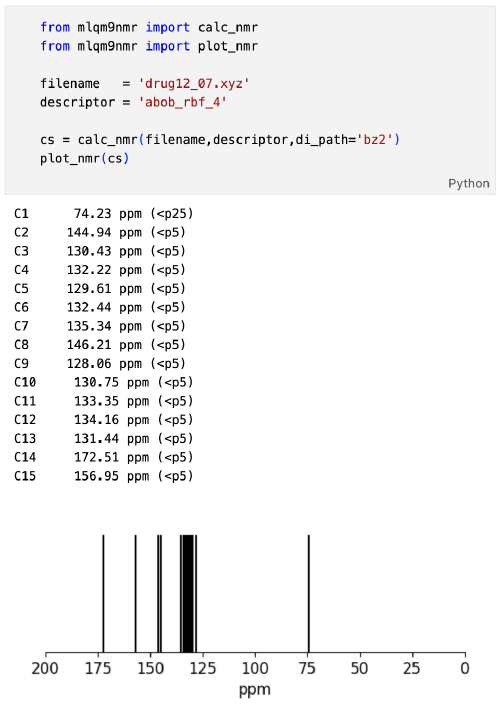}
    \caption{Example of predicting $^{13}$C-NMR chemical shifts using {\tt mlqm9nmr} module~\cite{mlqm9nmr}. For more detail, see {\tt mlqm9nmr} module in SI.}
    \label{fig:mlqm9nmr_main}
\end{figure}

\begin{figure*}[!htbp]
    \centering
    \includegraphics[width=\linewidth]{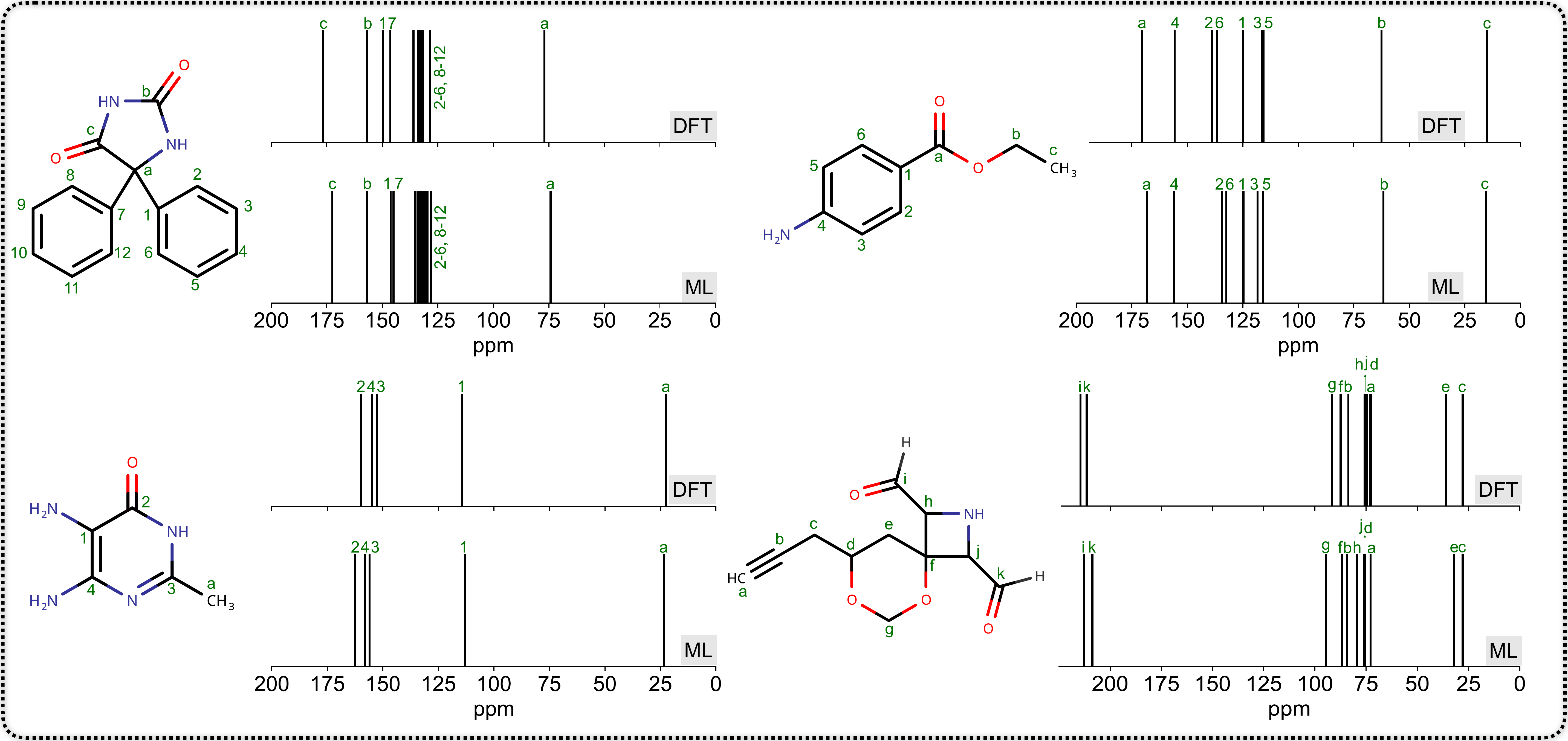}
    \caption{DFT calculated (NMR: mPW1PW91/6-311+G(2{\it d,p}), geometry: B3LYP/6-31G(2{\it df,p})) and ML-predicted  $^{13}$C-NMR spectra for randomly chosen molecules, each from Drug12, Drug40, pyrimidinone, and GDBm validation set. ML predictions were made with {\tt mlqm9nmr}~\cite{mlqm9nmr} using aBob-RBF(4) descriptor generated with PM7 geometries. See Figure~\ref{fig:mlqm9nmr_main} for an example query and SI for further details.}
    \label{fig:dft_ml}
\end{figure*}

\subsection{Transferability to external datasets of larger molecules}
        
To evaluate transferability, we first test the descriptors on larger drug-like molecules from the Drug12 and Drug40 datasets. As summarized in Table~\ref{tab:test_validation}, the predicted MAEs with aBoB-RBF(4) are lower than those obtained previously with FCHL~\cite{gupta2021revving} and 
MACE-OFF23-small~\cite{shiota2024universal}, showing that this descriptor can capture chemical environments not only for small QM9 molecules but also for larger systems. For the QM9NMR dataset, MAEs are significantly larger than the nearly vanishing mean signed errors (MSEs), indicating that the deviations from the DFT reference values are largely non-systematic and cancel out between positive and negative errors. By contrast, for Drug40 molecules (7--17 CONF atoms), the aBoB-RBF(4) descriptor exhibits slightly negative MSEs, suggesting a mild underestimation of chemical shifts as system size increases. This bias becomes more pronounced in Drug12 molecules (17--23 CONF atoms), where the MSE reaches about $-1$ ppm. Correcting for this systematic bias with a few reference values would lower the MAE to approximately 2.4 ppm. For comparison, the aSLATM(2) descriptor yields consistently small MSEs, reflecting a lower degree of systematic bias; however, its residual prediction errors remain non-systematic and thus less amenable to post-prediction calibration.

For the QM9NMR dataset, systematic improvements are observed for aSLATM($n$) as $n$ increases from 0 to 2, with both MAE and SDE decreasing slightly. Although aSLATM(2) achieves lower errors than aBoB-RBF(4) (MAE and SDE), its evaluation is computationally more demanding, despite only a modest increase in descriptor length (9315 vs.\ 8325). Importantly, for new datasets such as Drug40, aSLATM(2) performs slightly worse than aBoB-RBF(4), highlighting the latter’s better transferability.

We further validate performance on 208 pyrimidinone derivatives. Figure~\ref{fig:pyrimidinone_GDB_analysis}a shows the scatter plot of ML-predicted versus DFT-calculated $\delta(^{13}{\rm C})$ values, with an overall $R^2 = 0.996$ and an MAE of 2.19 ppm. The chemical shifts of $sp^2$ carbons are predicted with high accuracy ($R^2 = 0.978$), whereas predictions for $sp^3$ carbons are slightly degraded ($R^2 = 0.931$). While individual errors can be significant, the overall agreement across the dataset is excellent, facilitating large-scale NMR assignment.

Finally, we test 200 molecules from the GDBm dataset, a subset of the GDB17 database~\cite{ruddigkeit2012enumeration}. As shown in Figure~\ref{fig:pyrimidinone_GDB_analysis}b, predictions of $^{13}$C $\sigma_{\rm iso}$ values yield $R^2 = 0.995$ and an MAE of 2.66 ppm. As the number of heavy atoms $m$ increases, MAEs gradually rise (up to 3.06 ppm for $m=16$), and while $R^2$ remains at 0.99, consistent with the fact that the models were trained on molecules containing fewer CONF atoms.  

Overall, these results demonstrate that KRR models with the aBoB-RBF(4) descriptor not only achieve benchmark performance on QM9NMR but also transfer reliably to larger drug molecules, biomolecules, and GDBm molecules with more heavy atoms, making them broadly applicable for predicting $^{13}$C NMR chemical shifts.

\subsection{Example NMR prediction using the {\tt mlqm9nmr} module}
The 100k models developed in this study are distributed through an open-source Python package, {\tt mlqm9nmr}~\cite{mlqm9nmr}, which enables prediction of $^{13}$C chemical shifts for out-of-sample molecules. A usage example with the aBoB-RBF($4$) descriptor is shown in Figure~\ref{fig:mlqm9nmr_main}. Installation and usage instructions are provided in the SI. The package includes trained models for aCM(4), aCM-RBF(4), aBoB(4), and aBoB-RBF(4).

Quantifying prediction accuracy for a new query atom is not straightforward; however, it is useful to estimate whether the query lies within the chemical space represented in the training set. To assess this, the module employs a percentile-based metric. For a new query atom $q$, the minimum descriptor difference $D_{qt}$ (where $t$ spans all training atoms) is compared against the distribution of $D_{ij}$ values computed among the training set. If this minimum lies below the 5th percentile ($P_5$) of ${D_{ij}}$, the prediction is flagged as `<p5’, indicating that the query atom is well represented in the training data. Higher percentiles (such as
$P_{75}$ or $P_{95}$) suggest that the atomic environment is underrepresented, as may occur when applying the QM9NMR-trained KRR models to much larger molecules with distinct bonding patterns or conformations. Further details are provided in Figure~S4 and Table~S9 of the SI. The construction of training-set descriptors is illustrated in Figure~S5, and a step-by-step guide to predicting new molecules is presented in Figures~S6–S7.

Figure~\ref{fig:dft_ml} demonstrates ML-predicted $\delta(^{13}{\rm C})$ values (in ppm) for four molecules larger than those in QM9, using the aBoB-RBF(4) model. 
Across all four cases, ML predictions agree closely with DFT reference values, showcasing the applicability of these models for large-scale, quantitative NMR predictions in high-throughput screening and experimental structure assignment.

\section{Conclusions\label{sec:conclusion}}

In this study, we have formulated continuous versions of two discrete descriptors and evaluated their performance in predicting $^{13}$C-NMR $\sigma_{\rm iso}$ values. The local or atomic descriptors, aCM and aBoB, and their counterparts in the continuous domain, aCM-RBF and aBoB-RBF, show that the continuous descriptors can capture the structure-property relationship more effectively than their discrete versions.     
By convention, the atomic environment mapping of an atom is only done with the description of the target or parent atom. 

In this study, we have demonstrated that incorporating information from the nearest neighbors enhances the accuracy of the descriptors. With the augmentation of the descriptors of $n$ neighbors to the parent atom's descriptor, the improvement in the accuracy of the $^{13}$C chemical shift is gradual, but it saturates for $n=5$. As the maximum number of bonded neighbors for a C-atom is 4 ($sp^3$ hybridized), the choice of $n=4$ offers the least prediction errors.

The QM9 and QM9NMR datasets have been widely used as a benchmarking tool for ML architectures. Within the KRR framework, the aBoB-RBF($4$) descriptor achieved a low error of 1.69 ppm on the training dataset. While this accuracy approaches that of the target DFT reference, it is important to note that the DFT values were obtained using geometries optimized at the DFT level. In contrast, the ML models rely only on PM7 semi-empirical minimum-energy geometries. Since PM7 geometries are much faster to compute, this enables rapid, large-scale NMR predictions at near-DFT accuracy. However, starting from SMILES and optimizing with PM7 may yield geometries that differ from DFT-optimized structures, particularly in conformational preferences. This discrepancy likely contributes to the increasing prediction errors observed when applying QM9-trained models to larger molecules, where conformational landscapes become more complex and method-dependent.

To demonstrate transferability, the models were further validated on out-of-sample datasets, including Drug12, Drug40, pyrimidinone, and GDBm, achieving MAEs of 3.43, 2.80, 2.19, and 2.66 ppm, respectively. These results show that the ML models developed in this study are not only accurate for small QM9 molecules but also transferable to larger and more diverse chemical systems. 
Using the $\Delta$-ML approach~\cite{ramakrishnan2015big} with the aBoB-RBF(4) descriptor, the MAE for predicting 50k hold-out $^{13}$C values from the QM9NMR dataset at the target level mPW1PW91/6-311+G(2\textit{d,p}) drops to 1.16 ppm when trained on 100k values and using mPW1PW91/STO-3G as the baseline. 
Comparable errors were also observed when HF/STO-3G was used as the baseline.
This represents a substantial improvement over direct prediction of the target level (MAE, 1.69 ppm), bringing the error closer to that required for structure elucidation ($\approx1$ ppm). 
Hence the use of $\Delta$-ML
offers a promising route to enhance accuracy further and compensate for differences between semi-empirical and DFT-level geometries, thereby pushing performance closer to high-level theoretical benchmarks. It has not escaped our attention that the aBoB-RBF(4) descriptor, while demonstrated here for NMR shieldings, may find broader utility in modeling other atom-specific properties such as core-electron binding energies and atomic partial charges.

\section{Supplementary Information}
Supplementary information (SI) available:
1. Parameter optimization for continuous descriptors: 
a) Benchmarking angular grid (Table~S1),
b) Choice of widths of Gaussian RBFs and ABFs (Table~S2),
c) Choice of $R_{\rm max}$ for continuous descriptors (Figure~S1, Table~S3),
d) Choice of radial grid size (Figure~S2, Table~S4),
e) Choice of $\beta$ in scaling function (Table~S5),
f) Choice of damping function parameters (Table~S6),
2. $\tau$-plot (Figure~S3);
3. Descriptor vector sizes (Table~S7);
4. Choice of kernels (Table~S8);
5. Descriptor difference distributions (Figure~S4, Table~S9);
6. Tutorial on {\tt mlqm9nmr} module (Figure~S5--S7).

QM9 dataset is openly available at:
\url{https://github.com/raghurama123/qm9pack}.
QM9NMR dataset is openly available at:
\url{https://moldis-group.github.io/qm9nmr/}.
The dataset has been updated with the geometries and NMR shielding parameters of the pyrimidinone dataset. 
ML models trained on 100k QM9NMR $^{13}{\rm C}$ atoms are provided through the Python module \texttt{mlqm9nmr} along with a tutorial at: 
\url{https://github.com/moldis-group/mlqm9nmr.git}.

\section{Data Availability}
The data that support the findings of this study are within the article and its supplementary material.

\section{Acknowledgments}
We acknowledge the support of the 
Department of Atomic Energy, Government
of India, under Project Identification No.~RTI~4007. 
All calculations have been performed using the Helios computer cluster, 
which is an integral part of the MolDis 
Big Data facility, 
TIFR Hyderabad \href{http://moldis.tifrh.res.in}{(http://moldis.tifrh.res.in)}.

\section{Author Declarations}

\subsection{Author contributions}
\noindent 
{\bf SD}: 
Conceptualization (equal); 
Analysis (equal); 
Data collection (equal); 
Writing (equal).
{\bf RR}: Conceptualization (equal); 
Analysis (equal); 
Data collection (equal); 
Funding acquisition; 
Project administration and supervision; 
Resources; 
Writing (equal).

\subsection{Conflicts of Interest}
The authors have no conflicts of interest to disclose.

\section{References}
\bibliography{ref} 
\end{document}